\input harvmac

\noblackbox

\lref\strom{ A. Strominger, {\it ``Open P-branes,''} hepth/9512059} 
\lref\witcom{ E. Witten, {\it ``Some comments on string dynamics,''}
           hepth/9507121}  
\lref\Witcfth{ E. Witten, {\it ``On the conformal field theory 
       of the Higgs branch,''} Hepth/9707093. }  
\lref\abkss{O.  Aharony, M. Berkooz, S. Kachru, N.  Seiberg, 
 E. Silverstein, {\it ``Matrix description of interacting theories in 
6 dimensions,''} Hepth/9707079.  } 
\lref\abks{ O.  Aharony, M. Berkooz, S. Kachru,  
 E. Silverstein, {\it ``Matrix description of $(1,0)$ theories in 
 six  dimensions,''} Hepth/9707079} 
\lref\lowe{ D. Lowe, {\it  `` $E_8
  \times E_8$ small instantons in string theory,''} hepth/9709015.}
\lref\leeyi{ K. Lee. P. Yi, {\it ``Monopoles and instantons on partially 
    compactified D-branes,''} Hepth/9702107, Phys. Rev. D56(1997)3711-3717 }  
\lref\schper{M. Perry and J. Schwarz
{\it ``Interacting Chiral Gauge Fields in Six Dimensions and Born-Infeld
   Theory,''} Hep-th/9611065. }  
\lref\appsch{M. Aganagic, C. Popescu, 
J. Park,  J. Schwarz, {``Worldvolume action of the M theory 5-brane,''}   
                  hep-th/9701166,  Nucl.Phys. B496 (1997) 191-214} 
\lref\pt{  I. Bandos, K. Lechner, A. Nurmagambetov, P. Pasti, 
   D. Sorokin, M. Tonin
{\it `` Covariant Action for the Super-Five-Brane of M-Theory,''} 
 hep-th/9701149, Phys.Rev.Lett. 78 (1997) 4332-4334 } 
\lref\hlw{ P.Howe, N.Lambert, P.West. 
{ \it ``The self-dual string soliton,''}
hep-th/9710034}   
\lref\kal{ R. Kallosh, {\it ``Worldvolume supersymmetry,''}
              hep-th/9709069.  } 
\lref\seibsxt{ N. Seiberg, {\it ``Notes on theories with $16$ 
                             supercharges,''} hep-th/9705117. }  
\lref\abks{O.  Aharony, M. Berkooz, S. Kachru, E. Silverstein, 
       {\it ``Matrix theories for $(0,1)$ theories in 6 dimensions''}
    hep-th/9709118.    }
\lref\bss{ T. Banks, N. Seiberg, S. Shenker, 
           { \it ``Branes and Matrices,'' } hep-th/9612157, 
Nucl. Phys. B490 (1997) 91-106.  }
\lref\gkms{ D. Gross, I. Klebanov, A. Matytsin, A. Smilga, 
 {\it ``Screening v/s confinement in $1+1$ dimensions,''} 
hep-th/9511104, Nucl.Phys. B461 (1996) 109-130, }    
\lref\wati{ W. Taylor,  ``D-brane field theory on compact spaces,'' 
Hep-th/9611042, Phys. Lett. B394 (1997) 283-287. }  
\lref\susmot{ L. Motl, L. Susskind } 
\lref\seibdecoup{ N. Seiberg, {\it ``Matrix Description of M-theory
 on $T^5$  and $T^5/Z_2$, ''} hepth/9705221, 
Phys. Lett. B408 91997) 98-104.  }  
\lref\dia{D. Diaconescu, {\it ``D-branes, Monopoles and Nahm equations,''}
           hepth/9608163.  } 
\lref\hw{A. Hanany. E. Witten, {\it ``Type IIB Superstrings, BPS monopoles, 
 and three dimensional gauge dynamics,''} hepth/9611230, 
Nucl. Phys. B492 (1997) 152-190   }
\lref\Hitchin{N.  Hitchin, {\it ``The 
self-duality equations 
on a Riemann surface, ''} L. M. S.  55 (1987), p. 59    }
\lref\bending{E. Witten, {\it ``Solutions of 
 four-dimensional gauge theories
       via M theory,''} hep-th/9703166,  Nucl. Phys. B500 (1997) 3-42 }
\lref\bd{ M. Berkooz, M. Douglas, {\it ``Fivebranes in Matrix Theory.''}
           hepth/9610236, Phys. Lett. B395 (1997) 196-202.  } 
\lref\chs{ C.Callan, J.Harvey, A.Strominger }
\lref\setstern{ S. Sethi, M. Stern, {\it ``A comment on the
  spectrum of H-monopoles, ''}  Hep-th/9607145, 
  Phys. Lett. B398 (1997) 47-51.   }
\lref\dkps{ M. Douglas, D. Kabat, P. Pouliot, S. Shenker, 
            {\it ``D-branes and short distances in String Theory,''}
             hepth/9608024, Nucl. Phys. B485 (1997) 85-127 } 
\lref\gubig{ S. Gubser , I. Klebanov, {\it ``Absorption by branes and 
 Schwinger terms in the World-volume theory,''} Hep-th/9708005.  } 
\lref\maco{ A. Macocia {\it `` Metrics on Moduli 
                              Spaces of Instantons over Euclidean 4-space,''} 
          CMP135, 467-496, 1991.   }
\lref\susk{ L. Susskind, {\it ``Another conjecture about Matrix Theory,''} 
  Hep-th/9704080.  } 
\lref\vw{ C. Vafa, E. Witten, 
{\it``A strong coupling test of S-duality,''} hepth/9408074, 
   Nucl. Phys. B341 (1994) 3-77  }  
\lref\ztor{ Z. Guralnik, S. Ramgoolam} 
\lref\hacve{ F. Hacquebord, H. Verlinde }
\lref\vainst{ Vafa, {\it ``Instantons on D-branes,''}
               Nucl.Phys. B463 (1996) 435-442, hepth/9512078   }  
\lref\whn{B.  de Wit, J. Hoppe and  H. Nicolai, 
           { \it ``On the quantum mechanics of Supermembranes,''} 
Nucl. Phys. B305 (1988),  545-581. }  
\lref\bfss{ T. Banks, W. Fischler, S. Shenker, L. Susskind, 
             {\it ``M-Theory as a Matrix Model: A conjecture''} } 
\lref\grt{ O. Ganor, S. Ramgoolam, W. Taylor, 
             { \it ``Branes, Fluxes and Duality in Matrix Theory,''}
             hep-th/9612077, Nucl. Phys. B. 492, (1997) 191-204.  } 
\lref\seibwhy{ N. Seiberg, {\it ``Why  is the Matrix 
               Model correct ?''} hepth/9710009.  } 
\lref\senm{ A. Sen, {\it ``Zero branes on $T^d$ and Matrix Theory,''}
   hepth/9709220.   }
\lref\susskind{ L. Susskind, {\it ``T-duality in 
            M(atrix)  Theory and S-duality  in field theory,''}
              hepth/9611164 } 
\lref\ah{ M. Atiyah, N. Hitchin, {\it ``The geometry and dynamics of magnetic 
   monopoles,''} P.U.P 1988.     } 
\lref\nahm{ W. Nahm, {\it ``Self-dual Monopoles and Calorons,''} 
      Lecture Notes in Physics 201, 1984. } 
\lref\orset{ O. Ganor, S. Sethi, to appear} 
\lref\fmw{ R. Friedman, J. Morgan and   E. Witten, {\it ``Vector 
 bundles and F-theory, ''} Commun. Math. Phys. 187 (1997) 679-743 }  
\lref\gukov{ S. Gukov, {\it ``Seiberg-Witten solution from 
Matrix theory,''} Hepth/9709138.}
 \lref\hamo{ J. Harvey, G. Moore, {\it ``On algebras of BPS states,''} 
               hep-th 9609017}   
\lref\lt{ M. Luty, W. Taylor, 
{\it ``Varieties of vacua in classical 
             supersymmetric gauge theories,''} Hepth/9506098, 
                 Phys. Rev. D53 (1996) 3399-3405} 
\lref\setiib{ S. Sethi,  {\it ``The Matrix formulation  
                        of type IIB 5-branes,''} hepth/9710005.} 
\lref\plefwal{ J. Plefka, A. Waldron,
           {\it ``On the quantum mechanics of 
              M(atrix) Theory,''} hepth/9710104. } 
\lref\seibtalk{ N. Seiberg, talk at Rutgers. }  
\lref\dvv{ R. Dijkgraaf, E. Verlinde, H. Verlinde
            { \it ``Matrix String Theory,''} 
            Nucl.Phys. B500 (1997) 43-61, hep-th/9703030 } 
\lref\bs{T. Banks, N. Seiberg, { \it ``Strings from Matrices''}
            hep-th/9702187,   Nucl.Phys. B497 (1997) 41-55 } 
\lref\motl{ L. Motl, {\it ``Proposals on nonperturbative 
                  superstring interactions,'' } hep-th/9701025.  } 
\lref\dk{ S. Donaldson and P. Kronheimer, 
             {\it ``Geometry of four-manifolds,''} 
                     Clarendon Press, Oxford, 1990.} 
\lref\banks{ T. Banks, {\it ``Matrix Theory,''} hep-th/9710231. } 
\lref\vahag{ C. Vafa, 
     { \it ``Gas of D-Branes and Hagedorn Density of BPS States,''} 
          hep-th/9511088, Nucl.Phys. B463 (1996) 415-419. } 
\lref\witpq{ E. Witten, {\it ``New `gauge'  theories in six dimensions,''}
             hep-th/9710065.  } 
\lref\dvvbh{ R. Dijkgraaf, E. Verlinde, H. Verlinde, 
             { \it `` 5D Black holes and Matrix strings,''} 
          E-print Archive/Hep-th/9704018.    }
\lref\brodie{J. Brodie, 
{\it ``Two dimensional mirror symmetry from M theory,''} 
hep-th/9709228. } 
\lref\hanlif{ A. Hanany and G. Lifschytz, 
{\it ``M(atrix) Theory on $T^6$ and a m(atrix) 
Theory Description of KK Monopoles,''} hep-th/9708037. } 
\lref\dvvfve{ R. Dijkgraaf, E. Verlinde, H. Verlinde, 
     {``BPS quantization of the 5-brane,''} 
        Nucl.Phys. B486 (1997) 89-113, hep-th/9604055.  }

\line{\hfill PUPT-1736 }
\line{\hfill {\tt hep-th/yymmddd}}
\vskip 1cm

\Title{}{On Matrix models of M5 branes}

\centerline{$\quad$ { John  Brodie and Sanjaye Ramgoolam}}
\smallskip
\centerline{{\sl Joseph Henry Laboratories}}
\centerline{{\sl Princeton University}}
\centerline{{\sl Princeton, NJ 08544, U.S.A.}}
\centerline{{\tt brodie,ramgoola@puhep1.princeton.edu}}

\vskip .3in

 We  compare  the $(0,2)$ 
 theory of the single M5 brane decoupled from gravity 
 in the lightcone with transverse $R^4$,    
  and a  matrix model description in terms of quantum mechanics on 
 instanton moduli space. 
 We  give some tests  of the  Matrix model in the 
 case of multi fivebranes on $R^4$.  We extract constraints on 
 the operator content of the field theory of the 
 multi-fivebrane system by analyzing the  
 Matrix model. We also begin a study of compactifications 
 of the $(0,2)$  theory in this framework, arguing that 
 for large compactification scale the $(0,2)$ theory 
 is described by super-quantum mechanics on 
 appropriate instanton moduli spaces.


\Date{7/96}

\newsec{ Introduction.}

By  taking $k$ 5-branes
of M theory in $11$ dimensions approaching each  other 
with the eleven dimensional Planck length,  
$l_p$ going to zero, one finds an interacting
theory decoupled from gravity \strom\witcom. This theory  has a moduli space 
of vacua $(R^5)^k/S_k$, and it has $(0,2)$ supersymmetry
\seibsxt. How to  incorporate 5-branes in Matrix Theory \bfss\ 
was  approached in \bd\ by modifying the 
quantum mechanics of zero branes. Identification of the 5-brane
charge and a relation to instantons was established in \grt\bss\
and developed in the context of Matrix black  holes in \dvvbh. 
A concrete  proposal for describing the 
six dimensional theory of 5-branes by 
quantum mechanics on moduli space of instantons 
on $R^4$ was made in \abkss, and extended to IIA 5-branes in \Witcfth. 
We will mostly restrict the discussion to M5 branes. 
The $0+1$ dimensional quantum mechanics of
 \bd\ has  
$U(N)$ gauge symmetry, and   $k$ hypermultiplets 
in the fundamental of $U(N)$. Six dimensional Lorentz 
 invariance is expected at large $N$. Following
 \susk\ one expects that the finite $N$ theory describes 
 the $(0,2)$ theory  compactified with radius $R$  on a lightlike 
 direction $X^-$,   with momentum $P_- = N/R$.

As was discussed in \seibwhy\senm, M-theory compactified on a large 
light-like circle $R$ is related to $\tilde M$-theory compactified 
on a small space-like circle $R_{11}$ by a large boost in the 
eleventh direction with $\gamma = R/R_{11}$. The quantum mechanics
of $N$  heavy D0 branes in IIA string theory in the presence of 
$k$ 4-branes 
 is then mapped to 
the theory of 5-branes in M-theory 
compactified along a lightlike direction. 
Following \seibwhy\senm,  
we demand that $ {R\over \tilde l_p^2} ={  R_{ll} \over  l_p^2}$. 
The coupling constant of the quantum mechanics 
$g_{qm}^2 = { R_{11}^3 \over  l_p^6 }$.
 In the limit 
$l_{p} \rightarrow 0$, $R_{11} \rightarrow 0$ 
keeping ${ R_{11} \over l_p^2}$ fixed we have the quantum 
mechanics with hypermultiplets  of \bd.
 Distances in the transverse dimensions 
also must be rescaled ${ R_i\over l_p}  = {\tilde R_i\over  \tilde l_p} $. 
 Then we take the limit
$g_{qm} \rightarrow \infty$ 
( equivalently we  consider 
 energies  that are small compared to the energy $g_{qm}^{2/3}$ )
to decouple the Higgs branch from the
 Coulomb branch. For $k \ne 1$, quantum mechanics
on the Higgs branch contains the physics of the 
interacting  $(0,2)$  conformal theory \abkss.

    The $(0,2)$ theory of $k$ 5-branes of 
   M theory has a moduli space of vacua
   $(R^5)^k/S_k$. At the origin of the moduli 
   space, the theory is  superconformal 
     and has $U(k)$  gauge symmetry. The quantum mechanics
   on instanton moduli space describes the
   theory in the neighbourhood of the  superconformal 
   fixed point.

\subsec{ Some key points.}

 The emphasis in \abkss\ was on the interacting 
  theories, but similar arguments 
   can be used for the single 5-brane. 
  This will be our starting point. To set this up 
 we show how to define the decoupled 
 theory of a single 5-brane by starting from the
 action developed in \schper\pt\appsch.  
 
We will make a  careful identification of 
states obtained from the quantum mechanics 
with the states of a tensor multiplet,  
   using the $Spin(5)$ R-symmetry. This allows us 
to see that we have a multiplet of the  $(0,2)$ 
 theory in  six dimensions as opposes to say $(1,1)$. 

We present a way to deal 
with {\it super}-quantum mechanics on symmetric
products inspired by orbifold cohomologies. 
Essentially we extend beyond zero-energy states the 
prescription which works for zero energy. 
In the course of this discussion we will describe,  for $k=1$,  
the   construction   of non-zero 
energy states in the quantum mechanics, 
which, like the ground state, 
 are still annihilated by $8$ supersymmetries
( some non-linearly realized in the quantum mechanics).

 We will describe, for general $k$  how the structure of the 
 quantum mechanics on Higgs branch
allows us to see the decomposition 
of states into those associated 
with the $U(1)$ part and those associated with the 
$SU(k)$ part of the spacetime theory. This 
 decomposition arises from the fact that the 
Higgs branch naturally separates into strata. 
These strata  are easy to understand from the 
 equations describing the supersymmetric vacua 
of the 0-brane gauge theory. From the point of view 
 of the 4-brane theory they are related to subtle 
point-like instantons. Our treatment of super-quantum 
mechanics on this stratified space uses again 
the fact that superquantum mechanics is related to 
cohomologies.

We will study how group 
actions on the instanton moduli space,
again for general $k$,  
can be used to get information on  the 
symmetries and operators of the $(0,2)$ theory.  
In particular the action of $SU(k)/Z_k$ on the 
space of based instantons leads to the 
statement that the {\it local}  operators at the conformal 
fixed  point organize themselves into 
representations 
of the gauge group which have zero $Z_k$ charge. 

We give a short discussion 
of  toroidal compactifications. 
When the torus is large compared to all 
the scales in the correlation function 
of interest, then we may expect techniques
valid for the $(0,2)$ field theory in  
$R^4\times R^+ \times R^-$ to continue to be valid. 
We may thus expect that moduli spaces of 
instantons on $R^{4-d} \times T^d \times R^+ \times R^- $
to be relevant. We will present some evidence in favour 
of this conjecture,  starting  with a new
perspective on the derivation on the Matrix model of
 the $(0,2)$ theory. Then we discuss some aspects of
 the opposite limit
 of small compactification scale.

While work on this paper was being completed,  
a number of comments on related issues appeared in \banks. 
We have also learnt  that Matrix models for compactified  
$(0,2)$ theory have been considered  in  \orset. 
The  $(1,0)$ Matrix models have been discussed in \lowe\abks, 
and Matrix models for IIB  five-branes have been given 
in \witpq\brodie\setiib\hanlif.

\newsec{ On a single 5-brane of M theory} 
  There is an action given for the single fivebrane of M theory
   by \schper\appsch\pt, and the gauge fixing 
   required to obtain a $6$ dimensional $(0,2)$ field theory 
   is done in \kal.
   The theory of a single decoupled 5-brane 
   can  be defined to be this theory
   taken to the limit where 
 $l_p \rightarrow 0$. 

 The  bosonic terms of the action looks schematically 
 like :  
 \eqn\scac{ S= {1\over l_p^6}   \int d^6x ( 1 + H^2 + H^4 + \cdots ) }
Here $H$ is dimensionless.   
The factor $l_p^{-6}$ is needed in front 
 of the action because that is the tension of the 5-brane
 of M theory. 
  We get rid of the $l_p$ in front by redefining 
 the $H$. $H^{\prime}  = { H \over l_p^3} $. 
 It is easy to see that now the higher terms are suppressed
 as $l_p$ goes to zero 
\foot{ After we had developed this argument, 
  a very similar argument  for
 a decoupled single M5-brane being free was given by  N. Seiberg \seibtalk }. 
 The same argument 
 applies for the fermions, 
with the scaling $\theta \rightarrow l_p^3 \theta$, 
 showing that  the only surviving  terms in the 
 limit as $l_p \rightarrow 0$ are the fermion kinetic terms.
Terms of the form  
\eqn\supre{  ( \theta \Gamma \partial \theta ) H +
             ( \theta \Gamma \partial \theta )^2 ... } 
are suppressed in the $l_p \rightarrow 0$ limit. 
 This kind of rescaling is subtle if we were to consider 
 a theory with $U(1)^n$ gauge symmetry arising at a point of 
 broken symmetry in a non-abelian  theory \seibsxt, because 
 then a rescaling puts the coupling in the gauge transformations. 
 But since we are here dealing with a $U(1)$ theory only there 
 is no problem.

We will relate  this free tensor multiplet theory 
 to quantum mechanics on 
 the moduli space of $U(1) $ instantons. 
We will show that the correct quantum 
mechanics on instanton moduli space which follows from 
\bd\ is actually a free quantum mechanics, and we outline
 how to recover the spacetime theory of a free 
 tensor multiplet. Essentially the quantum mechanics gives
 a worldline formulation of the tensor multiplet theory
 in the light-cone gauge.

\newsec{ Remarks on the super-algebra. }

The action of \bd\ has $32$ supersymmetries
and a global $ Spin(5) \times SO(4)$ symmetry. 
The $SO(4)$ is 
 associated with  the $4$ directions 
 transverse to the light-like directions  and 
the $5$ directions 
transverse to the 5-brane. 
When the parameters  $x_0$ and $\theta_0$  
associated with motion on the Coulomb branch are set 
to zero, as appropriate for the description of the internal 
dynamics of the 5-branes, there are  $16$ supersymmetries left. 
They are   $Q_{I}^{ \alpha}$ and 
$\tilde Q_{I}^{\dot \alpha} $. The $(\alpha, \dot \alpha)$ 
are indices  in the fundamental of 
the  left and right $SU(2)$  in the decomposition 
 $SO(4)\equiv SU(2) \times SU(2) $. 
The index $I$ belongs to the spinor of $Spin(5)$. 
The $Q$ are linearly realised and the $\tilde Q$ are 
non-linearly realised.

 The above structure of the supersymmetries 
can be understood by noting that
they  are SUSY surviving the presence of the M5-brane 
( $4$-brane in IIA ), and that they split into 
two $SO(4)$ chiralities depending on whether 
they are preserved by the zero brane or not.  
This follows from inspection of the equations
\eqn\sus{\eqalign{ 
  Q_L  = - \Gamma^{0} \Gamma^{1}\Gamma^{2}\Gamma^{3}\Gamma^{4} \tilde Q_R \cr 
  Q_L  = \Gamma^{0} \tilde Q_R \cr }}
which give the supersymmetries preserved by the 4-brane 
and zero-brane respectively. 
Consider $C_4 =  \Gamma^{1}\Gamma^{2}\Gamma^{3}\Gamma^{4} $ 
 acting on $Q$. Combining the first equation in \sus\
with the condition  for supersymmetry  broken 
 by the zero branes we get $C_4 Q = -Q$.  Combining  it with 
 the condition for supersymmetry unbroken by the zero branes we get 
 $C_4 Q = Q$.
The presence of two chiralities under  $SO(4)$  is consistent 
with the  theory describing  the $(0,2)$  in the lightcone  
because the $SO(6)$ spinors decompose into $SO(4)$ spinors of
 both chiralities. 

Starting from the $(0,2)$  theory  we may write 
the superalgebra in a $Spin(5) \times SU(2) \times SU(2)$ 
covariant form
\eqn\supalg{\eqalign{ &  \{ Q_{I}^{\alpha} , Q_{J}^{\beta} \} = P_+ J_{IJ} 
              \epsilon^{\alpha \beta} \cr  
& \{ Q_{I}^{\alpha} , \tilde Q_{J}^{\dot \beta} \} = 
( \Gamma^A)^{\alpha \dot \beta}   P_A J_{IJ} \cr  
& \{ \tilde Q_{I}^{\dot \alpha} , \tilde Q_{J}^{\dot \beta}  \}
   = P_- J_{IJ} \epsilon^{\dot \alpha \dot \beta}
    \cr }}
$J$ and $\epsilon$ are the antisymmetric invariants of the
appropriate group. The $A$ index is an  index 
in the fundamental of $SO(4)$.  
A similar set equations is used in \seibdecoup . 
We have set central charges corresponding to extended
objects to zero, since we are mainly interested 
in the simplest background  of the $(0,2)$ SCFT in this paper. 
 We can convert the algebra to a form 
 where it is written as a set of independent 
creation and annihilation operators necessary to describe 
the tensor multiplet,  
in at least two ways. One keeps the symmetry 
$Spin(5) \times U(1) \times U(1)$. 
Another keeps manifest the symmetry 
 $SU(2) \times U(1) \times SU(2) \times SU(2)$, 
where the $Spin(5)$ has been broken to $SU(2) \times U(1)$.  
Either of these forms can be recovered from the
 Matrix model.  These will be  used in 
subsequent sections to describe the tensor multiplet. 
The superalgebra for the  Higgs branch Matrix model can be obtained
by starting  from the calculations of \bss, adding 
the contributions of the fundamental hypermultiplets, 
and specializing to the Higgs branch. The last step involves 
setting to zero all the scalars $X$ which transform under the
$Spin(5)$.

By inspection of the commutator of $Q$ with 
$\tilde Q$, we can identify the 
operators $P_I$ which generate translations
in $R^4$. The calculation is done in \bss\
in the absence of fundamental hypermultiplets, and it is easy to 
see that the  presence of these hypermultiplets does not modify 
the expression for $P_I$ \foot{ This form of the 
translation operator 
for the Matrix model of the $(0,2) $ theory has also been 
 considered   by O. Ganor.} : 
\eqn\opmom{  P_A = tr (\partial_t X_{A} ). }
This stems from the relation  
 \eqn\tildQ{ \tilde Q  = tr \lambda .  } 
which    involves the fermions
 in the adjoint  hypermultiplet but not those
in the vector.  
Suppose we are given a state  
in the quantum mechanics, which corresponds 
 to a {\it state }   of the $(0,2)$ 
theory  ${ \cal O}  (0) |0> $,  at a point $x^I =0$ in spacetime. 
We can find the state  corresponding to the 
operator at a generic point by acting 
 with $ e^{x^I P_I} $. 
\eqn\corrs{\eqalign{   { \cal O} (x^I ) |0> &=
            e^{x^I P_I} { \cal O}(0) |0> \cr }}
On { \it operators}, we have $e^{i x^I P_I}  { \cal O(0) }   e^{-i x^I P_I} $. 
Thus we can construct the transverse 
spatial dependence of states and operators. 
Similarly the Hamiltonian of the quantum mechanics
is $P_+$ of the  spacetime theory. This operator
does receive 
extra contributions  from  the hypers of the form : 
\eqn\pplus{ P_+ = |DH|^2 + |D\tilde H|^2 +  \dot \chi \chi . } 
In the special case $k=1$ which occupies the early 
sections  of this paper, $H= \tilde H = 0 $, and
these corrections to $P_+$ vanish. 

Note that $P_I$ has no dependence on the $H$ and $\tilde H$ 
fields. The space $M_{k,N}$ can be described by holomorphic
 gauge invariant coordinates, as discussed for example 
in  \lt. Coordinates made purely 
from the scalars of the  vector  hypermultiplet, of schematic form  
 $H\tilde H$ are inert under spacetime 
translation. 
One  combines the adjoint hypers into complex 
matrices: 
\eqn\compl{\eqalign{ &  U = X_1 + iX_2 \cr 
           &  V= X_3 + i X_4  \cr }}
Gauge invariant coordinates like 
$tr U^2$, $tr UHV\tilde H$  transform  non-trivially. 
So wavefuntions which are independent of $U$ and $V$ 
are inert under spacetime translation.

States in the quantum mechanics for $N$ instantons 
and with energy $E$ correspond to states in the 
$(0,2)$ SCFT with $P_+ = E$ and $P_-= N/R$:   
 \eqn\statecor{ { \cal O}_{ N,E} (0) |0> } 
By summing over instanton numbers 
we reconstruct states localized in 
  $x^-$ dependence, and by summing over $P^+$ we recover 
the $x^+$ dependence of correlators.

Many of the central charges  
that appear in \bss\ involve traces of
commutators  of matrices, so they vanish at finite
$N$. However at finite $N$ one can build 
matrices which approach the correct matrices 
at large $N$. 
For example in the construction 
of membranes we have $X_1= P$ 
and $ X_2 = Q$, where Q is diagonal and 
$P$ is a cyclic permutation. The construction
of such matrices when we are doing quantum mechanics
on the Higgs branch is restricted by the requirement 
that the constraints defining the vacuum 
 are satisfied. So for $k=1$, 
the $X$ are diagonal and $H=\tilde H=0$  \Witcfth. 
This is  an example of the  general fact that
while the central charges can be simply written down
as in \bss, new features can be expected from the 
specialization to the Higgs branch.

\newsec{ Instantons on $R^4$ and free tensor multiplet in $6$ dimensions.} 
The quantum mechanics  with gauge group 
$U(N)$ and $k$ hypermultiplets 
 with hypermultiplets given 
in \bd\ has a Higgs branch of vacua where 
 the adjoint or the fundamental  hypermultiplets
 acquire vevs. We will call this space $\bar M_{k,N}$.  
To derive the action for superquantum mechanics on 
 this space,  we give time dependence to the 
 coordinates on the moduli space, and 
 plug into the  action for the Higgs branch variables 
coming from \bd.  This standard 
procedure  for setting up 
the  collective coordinate quantum mechanics 
is used in this context in  \Witcfth.  
 The bosonic
 terms take the form : 
\eqn\formbos{ \int G_{ij}(Z) \partial_t Z^i \partial_t Z^j }

There is a 1-1 correspondence between the 
Higgs branch of the $U(N)$ gauge theory with 
$k$ flavors and the moduli space of $N$ instantons
in $U(k)$   gauge theory. Further, the metric in \formbos\
 is the same as the one on instanton moduli space  
\eqn\met{ G_{ij}(Z) = \int d^4 x \sqrt{g}
             \delta_{i} A^{\mu} (Z) \delta_j A^{\mu}(Z) } 
by a theorem of \maco.

\subsec{ Quantum Mechanics on $S^N (R^4)$ and free 6D theory.    } 

 In the simplest case $N=k=1$ 
the moduli space is just $R^4$ and the
action for the moduli space quantum mechanics 
 is just 
\eqn\act{ S = \int ( \partial X^I)^2 + fermions }
The $I$ index runs from $1$ to $4$, and the fermions
 will be discussed in detail in later sections. 
This a free quantum mechanics. All the interaction 
 terms vanish on the moduli space because they 
either involve commutators, or the hypermultiplets 
which are zero for $k=1$ \Witcfth. 
 
For $k= 1$ with general $N$, the moduli space
$M_{1,N}$ is the symmetric product
\eqn\symprod{  S^N (R^4) = (R^4 \times R^4  \cdots R^4)/S_N }    
This is the space of $N$ indistinguishable points
 on $R^4$. 
It  can be decomposed 
into a union of successively lower dimensional spaces  
as follows: 
\eqn\subspas{ 
S^N(X)  = \amalg_\nu (S^N(X) )_\nu }
where if $\nu$ is the partition
$(1)^{n_1} (2)^{n_2} \cdots (s)^{n_s}$
then
$$
(S^N( X ) )_\nu = \prod_i [ C_{X, n_i}]/S_{n_i}. 
$$
 $C_{X,n_i}/S_{n_i} $ is the space of  $n_i$ 
unlabeled separate  points on $X$. In this case $X = R^4$.

This space has orbifold singularities. 
String theory on such spaces 
is generally believed to be well 
 defined, and has recently been discussed 
for very similar orbifolds in the context 
of Matrix String theory \dvv\bs\motl. 
There the key point is that string theory 
on ${X^N \over S_N}$ has a Hilbert space 
which is that of $N$ strings.  
We will  argue that a very similar 
prescription works for superquantum 
mechanics.   
 A superparticle on the symmetric product 
of $X$, when the Lagrangian for motion on $X$ is free, will 
have states corresponding to the N-particle Hilbert space.

\subsec{ Spectrum} 
   The spectrum is trivial to solve in the case $k=N=1$. 
   The eigenstates of the 
  bosonic Hamiltonian  are parametrized by a 4-vector $k^I$. 
  Because of supersymmetry we   actually  have a supermultiplet 
   for each $k^I$.    
 For general $N$, we need to obtain the 
 spectrum for quantum mechanics on $S^N(R^4)$, 
 given the spectrum for quantum mechanics  
 on $R^4$. It is known how to relate  the 
 cohomology of $S^N(X)$ to that of $X$ \vw. Since
ground states of  super-quantum 
 mechanics  (SQM) are  typically  related to  cohomology, 
we have an obvious prescription for 
 obtaining the zero energy states 
for QM on $S^N(X)$. A simple guess, which 
we will argue is correct,   is that 
the   same prescription works for
 arbitrary states of the superquantum mechanics. 
 
The prescription   associates 
 oscillators $\alpha_{-l} ( h)$
to each cohomology class $h$, with $l > 0$. The 
cohomology of the symmetric product  is then  parametrized
by conjugacy classes of $S_N$. For a class associated
 with cycles of length $l_1,l_2, \cdots l_s$, 
with $l_1 + l_2 \cdots + l_s = N$,   
we have   states: 
\eqn\state{   a_{l_1}^{\dagger}(h_1)
a_{l_2}^{\dagger} (h_2) \cdots   a_{l_s}^{\dagger} (h_s)  |0> } 
In our case we will allow the state $h$ to be an 
arbitrary state of the quantum mechanics. 
So $h$ will be a supermultiplet labelled
 by a 4-vector $k^I$. As we will see the 
quantum mechanics contains non-zero momentum states
which are also BPS. So applying  the prescription to 
the case where the all the $h$ have the same transverse momentum 
is as well motivated as applying it to the ground  state. 
We will make the assumption that the obvious generalization 
of relaxing  this  constraint is correct, partly in analogy 
to the case of strings moving on these symmetric products
\dvv\bs\motl.

\subsec{ Oscillator number and  light-like momentum} 
 Our  interpretation of these states is that they 
   come from the free tensor multiplet in the six dimensional theory. 
  The subscript $l$ of the oscillator 
  can be interpreted as the  momentum in the 
  lightlike direction. To see this, consider 
   for example, a state of the form 
\eqn\sta{ a_{l_1}^{\dagger} (\vec k_1)  a_{l_2}^{\dagger} (\vec k_2)|0>.  }
  These are associated with subvarieties 
of $S^N (R^4)$ where the $N$ points form 
 two clumps of sizes $l_1$ and $l_2$, 
which add up to $N$. By specializing the 
 Higgs branch QM lagrangian to these configurations 
 we can calculate the energy to be 
$$  R \{ {  {\vec k_1}^2 \over l_1} + { {\vec k_2}^2\over l_2}   \} $$
as appropriate for a two particle massless state
in $6$ dimensions with transverse momenta 
$k_1,k_2$ and longitudinal momenta $l_1,l_2$. 
 More generally    the states with a fixed amount of lightcone 
   momentum in a free field theory are just given by 
    the appropriately symmetrized multiparticle states 
   with the total lightcone momentum partitioned between  
   the several oscillators.  
   But this is exactly how $N$ enters the Hilbert 
  space of our Higgs branch quantum mechanics.

\subsec{ Continuum of Normalizable States. }
 So far we have been discussing wavefunctions 
for  quantum mechanics on the Higgs branch of $U(N)$
 with $1$ flavour. This is 
appropriate for the infinite coupling limit, where 
Higgs and Coulomb branch decouple.   
 At finite coupling one has to do a quantum mechanics 
 involving the Coulomb branch.
 The wavefunctions for the 
 $(04) $  Higgs branch  states will spread over the Coulomb branch.  
Those that are BPS ( parametrized by a 4-vector
$k_A$ and a partition of
 $N$ ) will survive at generic coupling. 
So there should in fact be { \it a continuum of
 normalizable BPS states } on the Coulomb branch 
starting from the zero energy state studied
in \setstern\dkps.  Testing this by weak coupling 
methods would provide a partial 
test of the prescription we are using for doing SQM
 on these symmetric products.  
 We expect 
 that the same thing
 should be true for the case of 
  $U(N)$ gauge theory for $k$ flavours, and the density 
of states should be larger than the case of $k=1$. It 
should receive contributions from the $SU(k)$ degrees
of freedom of $(0,2)$ theory as well as the 
$U(1)$ part. More discussion on this decomposition is given 
in section 4.9.

\subsec{ Symmetries and the tensor multiplet. } 
 Here we review how  the symmetries of the 
 quantum mechanics on the Higgs branch 
 are related to those of the  $(0,2)$ theory
 in the lightcone. In particular we see
 how we distinguish  it from a $(1,1)$ theory 
  in a lightcone gauge. This allows us to unambiguously 
 identify the free particle described by the Higgs 
 branch superquantum mechanics  as a tensor 
 multiplet of $(0,2)$ as opposed to say 
 a vector multiplet of $(1,1)$. 

  The chiralities of the fermions surviving 
          the introduction of 4-brane  are not the same. 
         This is compatible with the fact that we are describing 
          a $(0,2)$ theory in a lightcone gauge 
          because a $4$  dimensional  $SO(6)$ spinor
          of definite chirality   decomposes under 
          $SO(4)$ into two spinors of both chiralities. 
         The chiralities of the spinors one gets by considering  
         a $(1,1)$ or a $(0,2)$ theory in a lightcone gauge 
         are the same. But the $(0,2)$ theory has a $Spin(5)$ 
         R symmetry which the $(1,1)$ theory does not have. 
         Since we are doing M5 branes as opposed to 
         IIA 5-branes, this symmetry remains manifest. If we 
         describe the  supermultiplets in a way 
        that keeps  this symmetry manifest, we can distinguish
         the tensor multiplet of the $(0,2)$ theory from 
         the vector multiplet of $(1,1)$ theory.

The action of the $(04)$ system as  described in \bd\ 
 has $32$ supersymmetries of which $8$ are linearly 
realised. When we set to zero the parameters
$x_0$  and $\theta_0$, there are only 
$16$ supersymmetries, of which $8$ are linearly realised. 
This is what we expect for the supersymmetries of a $(0,2)$ 
theory in the lightcone gauge. 
The  zero momentum state  preserves the  
all the linearly realised SUSY and 
breaks the non-linear ones. The non-zero momentum states
preserve some combination of linear and non-linear SUSY as discussed
 later.

We now describe in some  more detail 
the construction of the supermultiplet
of zero energy states. 
The broken SUSY generate the states of the supermultiplet. 
 It is convenient to perform 
 an $Spin(5)$ invariant quantization of the 
zero modes: 
We have fermionic  oscillators obeying: 
\eqn\fermos{ 
 \{  a_I ,  ( \hat a_J) \} = J_{IJ} } 
 $J_{IJ}$ is the spin(5) invariant tensor. 
They are built from the generators of the
kinematic supercharges of lightcone $(0,2)$ theory: 
\eqn\build{ 
 \{ \tilde Q_I^{\dot \alpha} , \tilde Q_{J}^{\dot \beta} \} = 
\epsilon^{\dot \alpha \dot \beta}J_{IJ} 
} 
as follows: 
\eqn\osc{\eqalign{ &  a_{I} = \tilde Q_{I}^{1} + \tilde Q_{I}^{2} \cr 
                   &  \hat a_{I}
                       = \tilde Q_{I}^2 - \tilde Q_{I}^{1} \cr }}
 In some basis $J$  has components : 
$$J_{12} = -J_{21} = J_{34} = - J_{43} = 1.  $$ 
A subtlety is that we do not directly 
see the  $ SO(5) \times SU(2) $ covariant form of 
the algebra of broken supercharges \build\ 
in the Higgs branch quantum mechanics. But we do see 
the $SO(5) \times U(1)$ covariant form \fermos.  This is related to the 
fact that in the pure zero brane system, one has because
 of the Majorana nature of the fermions, to perform 
 an $SO(7)\times U(1)$ covariant quantization 
as opposed to a full $SO(9)$ covariant quantization \whn\plefwal.

 We can define operators transforming in the adjoint
 of the $Spin(5)$, since the   symmetric tensor product 
 of the spinor of $SO(5)$ is the adjoint, by 
\eqn\adj{ O_{IJ} = 1/2 ( {\hat a_{I} }a_{J} + {\hat a}_{J} a_{I}) }  
 A representation  of the superalgebra  is constructed 
by acting with  the 
 creation operators on the state $|0>$. 
Clearly $|0>$ is annihilated by $Spin(5)$. 
  The state build by four creation operators 
$\hat a$ acting  on the vacuum  is 
 invariant. 
$J_{IJ} ( \hat a_{I})  (\hat a_{J}) |0>$ is also invariant.
 The operators quadratic in the oscillators  decompose 
 into the  vector and the trivial rep of $SO(5)$ because
 the antisymmetric tensor product of the two spinors 
 decomposes as  $5+1$. 
We have, therefore, one vector of $Spin(5)$ and three singlets. 
This corresponds to the 5 scalars and the three polarizations 
of the antiself-dual tensor. 
Without analyzing the $Spin(5)$ content of the 
 states we cannot tell if our theory has something to do with 
 the $(0,2)$ in six dimensions or the $(1,1)$ in six dimensions, 
because each $SO(6)$ spinor decomposes into spinors
 of $SO(4)$ of both chiralities. However we are  able
 to distinguish a tensor multiplet in the lightcone frame
from a vector multiplet in the lightcone frame, as long as
 the momentum of particle is chosen to be entirely 
 in the plane chosen to be parametrized by the light-like
coordinates in going to  lightcone gauge. 
 The $5+3$ split tells us that we have a $(0,2)$ algebra
 tensor multiplet in 6D as opposed to a $(1,1)$ algebra. 

We digress to 
remark on the case of IIB 5-branes \witpq\setiib. 
There the proposed Matrix model comes from  
 zero branes moving on ALE space. The Coulomb branch 
SQM describes the 5-brane. In that case  the SQM has an 
  $SO(4)$ which is identified with R symmetry in the 
 5-brane worldvolume. And the analogous discussion 
 above will give 4+4 split of the bosons, 
as appropriate for a vector 
 multiplet in six dimensions, where the gauge field has a 
$4$ components and there are $4$ scalars.

It has been convenient in this discussion 
to perform a quantization with $Spin(5)$ 
covariance here. We could also 
quantize by keeping the $SO(4)$ parallel 
to the 4-brane manifest. Then we have to break the 
$Spin(5)$ to $SU(2)\times U(1)$. But the relation between 
the zero momentum and the non-zero momentum states 
is then clearer.

\subsec{ States carrying momentum in the quantum mechanics} 

   When we turn on some momentum we expect 
   that the SUSY left unbroken by the state 
    is some linear combination of $Q_{\alpha}^{ \rho} $ 
     and $\tilde Q_{\alpha}^{\dot \rho}$, which were interpreted
     as the SUSY of the worldvolume theory of the 5-brane. 
In fact we can see, in the quantum mechanics 
on Higgs branch,  that there are non-static configurations
 corresponding to eigenstates of momentum which are SUSY. 
In the simplest case we just have $U(1)$ with instanton number 
 $1$ and we have QM on $R^4$. The configuration   
\eqn\config{    X_{ \mu} = v_{\mu} t,   } 
The variation of the gluino vanishes
 for an appropriate combination of linear and non-linear 
 SUSY:  
\eqn\var{ \delta \lambda = \Gamma^{\mu}  v_{\mu}  \epsilon + \tilde \epsilon } 

 We can also see this using the  lightcone superalgebra 
 acting via a more quantum mechanical construction. 
 We consider the operator 
\eqn\optor{ e^{ik_A  X^A } \Psi_k,  }
where $\Psi$ is the fermionic part.  
For $k=0$ the fermionic vacuum is built 
 from the broken SUSY $\tilde Q^{\dot \rho}_{I}$, and 
is annihilated by all the linear SUSY $Q_{I}^{\rho} $.
This is consistent with the relation from the 
superalgebra 
\eqn\supera{\eqalign{ &   \{ Q,Q \} = P_{+} = 0. \cr 
       &        \{ Q, \tilde Q \} = 0  \cr } } 
For non-zero $k_A$ we insert into the lightcone superalgebra
of the  $(0,2)$ theory,  
\eqn\supnonz{\eqalign{ &  P^+ = R k^2/N \cr
                       &  P_A = k^A \cr 
                       & P_- = 1 \cr   }}  
( we have set $R=1$ ) to  find that
$k_{\mu} \Gamma^{\mu}  Q + \tilde Q$ 
can be consistently set to zero. The combination 
with the opposite sign can be separated 
into creation and annihilation operators 
to build a representation of the superalgebra. 
Requiring  a correct representation of the 
superalgebra with $k \ne 0$ fixes  the fermionic
part to be of the form given below : 
\eqn\ferm{ e^{i k_{A}(  X^{A}  + 
b^i_{ \alpha } ( \Gamma^{A})_{\alpha \dot \beta  }
b^{\dagger}_{i \dot \beta} )  } 
  |0> }
Here $b$ are linear combinations of the 
fermions  $\lambda_{I}^{\dot \alpha}$  which appear in the 
SUSY quantum mechanics action. The lower case $i$ index 
is in the fundamental  of an $SU(2)$ subgroup 
of the $Spin(5)$.  
The state  $|0>$ is an $SO(4)$ invariant vacuum, 
annihilated by $b$.  It is  thus possible to maintain the  
$SO(4)$ covariance at the quantum mechanical 
level, when we give up the $Spin(5)$ covariance.

\subsec{ Propagator } 

The generic state  in this SQM on  $ (R^4)^N/S_N$ 
in \state\ is a multi-particle state. 
The $\vec k_i$ are momenta which characterize the states
 of the QM in $R^4$. 
To make the identification with one-particle states
of the tensor multiplet, we identify 
\eqn\cores{ \partial_{- }  B (n, \vec k ) |0>  \rightarrow 
a^{\dagger}_{n} ( \vec k ) |0>  } 
By simply taking inner products in the 
 quantum mechanics we can get the two point function: 
\eqn\twopt{ 
  \langle \partial_{-} B(x^{-}, x^{i} )  \partial_{-} 
B ( y^{-}, y^{i} \rangle  = \delta^{\prime}  
(x^{-} - y^{-} ) \delta ( x^{i} - y^{i} ) } 
The field $B$ can be thought as one of the components 
 of the self-dual tensor.

Recall  that the actual gauge fixing to lightcone
 gauge may be subtle in the
 action given by \pt\appsch\ since,  for the choice 
$a= x^{-}$, we have  a singular action.
Here $a$ is the auxiliary field which enters the 
Lorentz covariant action of 
 The  matrix 
model gives a 
  gauge fixed, worldline formulation  of the 
free tensor multiplet theory in the light-cone. 

\subsec{ Correlators } 
The inner product if the field theory 
can be mapped in our $k=1$ example
to the natural  inner product in the quantum 
mechanics
\eqn\inprod{ \int \sqrt{ G(Z)} \Psi^* (Z) \Psi (Z) } 
The obvious generalization is to integrate 
products of several wavefunctions which 
should correspond to correlators in the
field theory.

\subsec{ Subvarieties of  the Higgs Branch. }

There is a decomposition of the Higgs branch $\bar M_{k,N}$ : 
 \eqn\dec{ \bar M_{k,N} = M_{k,N} ~  \amalg ~ M_{ k,N-1} \times R^4  \amalg 
 \cdots M_{k,1} \times S^{N-1}R^{4} \amalg S^N (R^4) }
$\bar M_{k,N}$ is a connected space. 
This is Higgs mechanism in the zero-brane worldvolume
theory, but is {\it not}  Higgs mechanism in the spacetime
$(0,2)$ theory. (The latter would correspond to giving masses
to the fundamental hypermultiplets). 
 The component containing $M_{k,N-l}$ 
corresponds to the subspace where we have set $l$ components of 
$H$ and $\tilde H$ to zero.   This decomposition is also 
studied in the context of instantons \dk. 
The component with the $l$'th power of $R^4$ corresponds to the 
subspace where $l$ instantons are pointlike and 
$N-l$ are fat.

The fact that for a given $l$,   the symmetric product part 
has no dependence on $k$ contains an important property 
of pointlike instantons : they do not carry gauge indices, 
and cannot be described by a potential or field strength.  
This will turn out to be just right for the physical 
interpretation of this decomposition, 

We will argue that  the wavefunction
for 
$ \bar M_{k,N}$ 
 splits into  a sum over each component. 
The last factor would describe the decoupled centre of mass motion  
described by a free tensor multiplet. 
More  generally  the decomposition can  be interpreted 
 by saying that of the $N$ units of momentum, 
 $l$ are carried by the interacting part ( $SU(k)$ )  of the 
 theory, and $(k-l)$ are carried by the 
free part ( $U(1)$  )  of the theory.
\eqn\decomp{ 
  H(k,N) = \oplus_l | s_1(l) > \otimes | s_2(N-l) >
} 
where $|s_1>$ is  a state in the $SU(k)$ part of the 
theory, and $|s_2>$ is a state in the $U(1)$ part of the 
theory.

The fact that this is the correct way to deal with 
the disjoint union can be motivated as follows. 
Since we are dealing with supersymmetric quantum 
mechanics, we have a close relation to the 
cohomology of the space. At the level of cohomology, 
the correct prescription for the disjoint union  
 is certainly to  take the direct sum of states 
from each component. This   follows from  
 the Mayer-Vietoris sequence for the appropriate cohomology. 
We have seen that there are also non-zero energy states
which can be related to the zero energy states 
by transformations which have the interpretation of 
boosts parallel to the 4-brane worldvolume, and which also preserve
$8$ supersymmetries.
 So certainly for this class of states 
taking the direct sum of the states obtained from 
 each component is the correct prescription. 
 It is plausible,   then,  that 
the same prescription applies to  the entire 
 super-quantum mechanics.

 The very important property that the
 pointlike instantons 
 do not have a  connection and field strength 
 is crucial  here. We can decompose the 
$tr F^2$ in terms of contributions from 
pointlike instantons and fat instantons \maco\ 
but not the connection or field strength.
This is related to the fact that 
 we need generalize the concept 
of bundle to sheaves when dealing with 
point-like instantons,  which  has been emphasized 
 in a related context in \hamo.  
If it were possible to associate some internal 
`gauge indices' to the point-like instantons, we would 
expect some dependence of the spaces of point-like instantons
on $k$. So  we would have, say $(R^4)^{Nk}/S_{Nk}$,  
which could {\it not }   be interpreted in terms of 
 a free tensor multiplet theory describing the centre
of mass at momentum $N$. 
 
In the above we have not been too precise about 
exactly what kind of cohomology corresponds to the 
ground states of the quantum mechanics. 
Clarifying this will  be very interesting, for example
the  states coming from the open subset  
 $M_{k,N}$ will be constrained by the  uniqueness of the 
 vacuum of the $SU(k)$ theory.

\newsec{  Group Actions on Moduli spaces.  } 
 In doing quantum mechanics on a space 
 parametrized by $Z$, if there 
 is an action of a group $Z \rightarrow g Z$, then there 
is also an action on wavefunctions
\eqn\wavefns{ \Psi ( Z )  \rightarrow \Psi (g Z) } 
If $g$ acts trivially,  then the wavefunctions 
transform trivially. 

  We will see in this section how 
  some conformal symmetries of 
  a 6D  conformal theory lead to manifest 
  symmetries of its lightcone formulation 
  and see how these symmetries are realized
  in the proposed SQM on instanton moduli space. 
  
 Then we will see how the action of the
 global gauge transformations 
 on instanton moduli spaces lead to some 
 plausible constraints on the {\it local }  operator content 
  of the $(0,2)$ theory.

\subsec{ Spacetime symmetries.} 

For a six dimensional theory in the lightcone
we expect a  manifest $SO(4)$ group  of rotation symmetries, 
as well as a group of translations. The $SO(4) $ appears as 
an $R$ symmetry of the quantum mechanics. The generators of
the translation group were constructed in section 3.  

Now we consider the conformal invariance of
the six dimensional theory.  
Consider, first, the  scale transformations
 of the $(0,2)$ theory. Thay act as  
 \eqn\sca{\eqalign{ &  x^+ \rightarrow \lambda x^+ \cr 
                    &  x^- \rightarrow \lambda x^- \cr 
                    & x^i \rightarrow \lambda x^i,  \cr }} 
where $x^{\pm} =  x^0 \pm  x^5$ and $i$ runs from 
$1$ to $4$. 
A Lorentz transformation can be done  
 to get rid of the change in $x^{-}$. 
This is a desirable  thing  to do 
because symmetries which act on  $x^-$ can only 
be seen when we reconstruct the $x^-$ dependence 
of correlation functions by summing over instanton numbers. 
 This means that we have
\eqn\scalo{\eqalign{ &  x^+ \rightarrow \lambda^2  x^+ \cr 
                    &  x^- \rightarrow   x^- \cr 
                    & x^i \rightarrow \lambda x^i \cr 
}}  
giving us transformations of the momenta: 
\eqn\momtrns{\eqalign{ & P_+ \rightarrow \lambda^{-2} P_+ \cr 
                       & P_- \rightarrow P_{-} \cr 
                       & P_i \rightarrow \lambda^{-1} P_i \cr }}
We can check that these transformations are indeed obeyed by the 
 simplest moduli spaces ($k=1$, any $N$)  which are  symmetric products
of $R^4$ with the metric inherited from the Euclidean metric 
 on $R^4$. 
For more general cases it seems to put 
 a conformal invariance requirement  on the metric
of instanton moduli space \met.

\eqn\metconf{ G_{\mu \nu} ( \lambda x ) = G_{\mu \nu }  ( x) }  

  We can see
 this directly from the general definition 
 of the metric 
\eqn\met{ G_{ij} = \int d^4 x \sqrt{g} \delta_{i} A^{\mu} \delta_j A^{\mu} } 
 This definition  suffices to prove it is 
 conformal. When the coordinates of the instanton moduli space
 transform, we can find transformations of $x$, $ A(x)$ which  leave 
 the metric invariant: 
\eqn\trnsinst{\eqalign{  
&  Z_i \rightarrow \lambda Z_i \cr 
&  x \rightarrow \lambda x  \cr
& A(\lambda x , \lambda Z )  = \lambda^{-1} A(x, Z) \cr }} 
This is also seen from the
quantum mechanics, by noting that, 
in units where the $X$ and $H$ have worldline dimension 
$1$, all the terms in the action for the Higgs branch 
quantum mechanics are of same dimension. 

We have seen then that  the conformal 
invariance in $6$ dimensions
 has a simple implication for the
form of the quantum mechanics. 
We might  have expected to recover in a simple way 
 the special conformal transformations 
which  form part of the conformal 
group of $R^4$,  (labelled by a 4-vector 
$b^I$) 
but that  turns out not to be true.   
The special  conformal transformations,
 have an action which mixes the 
$x^-$ ( the hidden dimension 
dual to the instanton number ), with 
the $x^i$. This action is of course expected 
to act on the correlation functions we reconstruct 
by summing over instanton number, but is not 
a simple symmetry before summing 
over instantons. 
Interestingly, \maco\ finds in relating  group actions 
on  ADHM data to group actions on  
instanton moduli space data, that the dilatation 
subgroup of the conformal group of $R^4$ is distinguished
from the special conformal transformations.

\subsec{ Gauge group action}  
The correct instanton moduli space corresponding to 
 the Higgs branch  quantum mechanics
is that of based instantons \Witcfth, where two instantons related by a
gauge transformation which is not the identity at infinity 
are considered inequivalent. Without this definition, 
the space does not have a dimension which is a multiple 
of $4$,   as needed if it is to admit  SQM with so $8$ SUSY.
 Based  instantons are considered gauge 
inequivalent if they are related by a global
gauge transformation. 
 This means that the moduli space 
has an action  of $SU(k)$. 
The $U(1)$ part of $U(k)$ acts trivially 
because it commutes with all the fields 
entering the instanton solution, which are in the adjoint. 
It follows that the centre $Z_k$ of the $SU(k)$ part also acts 
trivially. 
 This means that  
  wavefunctions will form representations
   of $SU(k)/Z_k$  for a theory  
  of $k$ 5-branes.
Using the operator-states correspondence
this leads us to the prediction that 
all the local operators of the 
$(0,2)$ transform under the  $SU(k)$  
gauge group of the interacting  theory as 
representations which carry zero $Z_k$ charge. 
This includes for example states which are in the adjoint
or its tensor products but not the fundamental. 

 This may at first seem surprising from the
point of view of the 
    zero brane worldvolume theory, since it contains
fundamentals, $H$, $\tilde H$ under the $U(k)$ flavour symmetry. 
However, the vacua may be parametrized only by 
combinations of  these variables which are   invariant under the 
$U(N)$ gauge symmetry. These  
are representations of   $U(k)$ with 
zero $Z_k$ charge. Indeed, the  gauge invariants are built
 from polynomials in the $U,V,H,\tilde H$. Only the 
$H$, $\tilde H$  carry $U(k)$ indices. The $U, V$ have two 
$U(N)$ indices, so to form a gauge invariant
quantity the total number of $H$ and $\tilde H$ 
is even,  which means that the gauge invariant object 
cannot transform as a fundamental. 
   
This constraint on the operator
content of the $(0,2)$ theory   
 is not in contradiction with 
the fact  that one can find a string soliton  \hlw\
ending 
 on a single 5-brane, which is charged under the 
$U(1)$.   The soliton is {\it not} 
 created by a { \it local operator} 
in the field theory.   The set of states obtained from the 
quantum mechanics is expected to be related to the spectrum 
of local operators acting on the vacuum. While the algebra 
of local operators only contains  adjoint representations 
the non-local operators can transform in the fundamental 
representation. Analogous phenomena in two-dimensional field 
theories are known \gkms.

   The  constraint on the local operator
content is  plausible because as we move away from 
   the origin we expect adjoint objects like strings. 
   But this is a stronger statement, since it is 
   a statement about  the origin of moduli space. 
  In the context of  the $4$ dimensional $N=4$ theory, 
  which is obtained by the dimensional reduction 
  of the $(0,2)$ theory, this is quite plausible.   
  We can consider operators obtained by taking composites 
    of the fields that enter the action of   $N=4$ 
 Yang Mills, and ask how their correlators
  behave as we approach the fixed point. This way we have  
  objects that transform in the adjoint and its tensor 
 products.     In the six  dimensional case we do not have 
   an action where the strings appear  as  adjoint fields 
  so we cannot use this   argument.

\newsec{  Compactifications of the
        $(0,2)$ theory }

We consider compactifications
 of the $(0,2)$ superconformal theory on a torus 
$T^d$ with the sides of the torus 
being of the same order of magnitude. 
We can consider the 
dependence of the  correlation functions 
on the spacetime coordinates
 and the compactification scale $L$: 
\eqn\cor{  < {\cal O} (x_1) {\cal O} (x_2) \cdots { \cal O} (x_k) >_L   
              = f(x_i,L) }
If we take $L$ to be large 
compared to $|x_i-x_j|$, for all $i,j$ we expect 
an asymptotic expansion of the correlation functions
 to exist with leading term given by the flat space 
correlation functions. 
It seems  reasonable to conjecture that this large
$L$ expansion can be reconstructed by considering 
quantum mechanics on instanton moduli space
on $R^{4-d} \times T^d$.
We will present
some arguments in favour of 
this.

The approach of \abkss\ 
in deriving the conjecture that 
the Matrix theory of 5-branes
is given by quantum mechanics on instanton moduli space starts with the 
worldvolume theory of  the  zero branes. An alternative approach 
which may 
 be trusted for instantons of finite size is the following.
 We start with the  $(0,2)$ field theory compactified 
on  a small circle of radius $R_5$, 
This is weakly coupled $4+1$ Yang Mills theory. We could 
use $1/R_5$ as an  ultraviolet  cutoff.  
We want to  look at the sector with 
momentum  along the $5$ direction. 
These appear as solitons in $5$ dimensions, 
obtained by embedding the instantons 
of $4D$ gauge theory into $4+1$ Yang Mills. 
Since $R_5$ is small, these are very heavy. Their 
non-relativistic dynamics is governed 
 by a quantum mechanics on the moduli space of the solitons. The simplest 
quantum mechanics is the supersymmetrization of the 
action of the form \formbos. 
There can be higher derivative terms but they should be suppressed by 
powers of $R_5$, since this is the parameter 
which measures the strength of quantum corrections. 
So the QM action is indeed the simplest  one. 
Now following \seibwhy\ we argue that the $(02)$ 
theory compactified on a lightlike circle 
is related by a boost to the $(0,2)$  theory 
on a very small spatial circle. How we treat the pointlike
instantons is not easy to motivate from this point 
of view, but at least  one   consistent way to do it
is to mimic the SQM on the Higgs branch of the  model of \bd.
It might be interesting to see if the symmetries of the 
problem could be used to constrain, from the $4+1$ 
dimensional point of view,  the quantum mechanics
to be exactly that of the Higgs branch. 
The advantage of developing this line of argument, 
is that it starts directly with the decoupled
 theory as opposed to 5-branes embedded in 
M theory. Another attraction of this approach is that 
it would elevate  the ADHM equivalence between self-duality 
equations in 4 dimensions and  matrix equations in $0$ 
dimension, to a quantum equivalence between two descriptions 
 of $(0,2)$ in the lightcone gauge.

Now we can put the $4+1$ Yang Mills 
theory on a manifold $R^{4-d} \times T^d$. 
The theory still has solitons which  are 
obtained from embedding instantons.
For large compactification scale, $L_c$, 
$4+1$  Yang Mills is valid.  Again the quantum corrections are suppressed
by powers of $R_5$,  so  we can trust 
the minimal quantum mechanics on the moduli space 
of instantons on $R^{4-d} \times T^{d}$.
Upon compactification of the $(0,2)$ theory 
there will arise many sectors due 
to Wilson surfaces of the two-form field, 
and fluxes. We will not attempt to 
give a comprehensive
discussion of all these sectors.

For generic compactification size
we would expect a description in terms of a
  $d$ dimensional field theory
as the Matrix model \wati.  This d-dimensional 
theory is by construction a field theory obtained by 
T-dualizing the  zero brane worldvolume theory.
We will not go into the detailed construction 
of this theory (which is developed in \setiib\orset ) 
but we will make some general remarks on its symmetries
and dynamics, based on what we expect from the 
properties of compactified $(0,2)$ theory. We will also 
discuss qualitative aspects of
the dynamics as  viewed from the 
$(4-d)$ dimensional brane worldvolume theory.

\subsec{$S^1$ compactification }

Instantons  satisfy  $F = *F$. 
Instantons on $R^3 \times S^1$ can be constructed
by an analog of the ADHM construction \nahm. 
Typically they have a non-trivial 
dependence on the coordinate living on the 
$S^1$ ( the ``calorons'' of \nahm ).
The ADHMN construction  of
the calorons is very similar to that 
of  monopoles. 
These are special instantons which have 
no dependence on the $S^1$ of $R^3 \times S^1$. 
They  obey the dimensionally reduced form 
of the self-dual Yang Mills equations.  
$$ F = D\phi, $$
 where $\phi = A_4$. 
The metric on the moduli space of monopoles is 
hyperkahler so it  admits the extended SUSY 
QM with $Spin(5)$ symmetry  that we want for the description 
of the system at large compactification scales \ah.
 The $Spin(5)$ is a consequence of hyperkahler 
geometry the way $SU(2)$ is a consequence of Kahler 
geometry \hamo. 
Moduli spaces of calorons are relevant to 
the $S^1$ compactified $(0,2)$ theory in the vacuum sector.
 The moduli spaces
of monopoles would be  relevant to the sector of DLCQ  $(0,2)$ theory 
on an $S^1$, with a Wilson surface $B_{4-}$

With the $S^1$  compactification, 
we can perform a T-duality 
converting the  D0 brane-D4 brane ( with 
spatial extent in  $(x^1,x^2,x^3,x^4) $ )  
system to D1 brane (extended along the compact direction 
$x^4$ )  intersecting 
D3 branes on with spatial extent $(x^1,x^2,x^3)$. This 
 system has been studied in connection with  monopoles 
in the 3-brane worldvolume\dia, \hw.
Now we  also have calorons 
which  correspond to D-strings that wrap the circle
an integral number of times \leeyi.

The (0,2) theory on $\bf{R^3 \times S^1 \times R^{+} \times R^{-}}$
is 4+1 SYM in the lightcone frame and 
is described by quantum mechanics on caloron  moduli space. 
By compactifying the $(0,2)$ theory on a circle of small radius $R$,
the coupling constant for the 4+1 SYM is $1/g_4^2 = 1/R_4$
which introduces a scale. The compactified $(0,2)$ theory is no longer 
scale invariant. In  the opposite limit of
small  radii,
we do not expect  quantum mechanics on moduli space 
to be valid. In this limit the compactification 
radius of the $(0,2)$ theory is small. 
Equivalently we are looking, in the
$4+1$ dimensional theory, at energies small 
compared to the scale set by the compactification 
radius.  The 4+1 SYM theory flows to a non-interacting 
theory at long wavelengths since $g_4^2$ looks very small 
compared to the scale at which we are doing QFT. 
In the 
Matrix description on the $1+1$ worldvolume, 
this limit should 
be related to an equivalent $1+1$ dimensional theory
flowing to a free fixed point.

Momentum in the $x^4$ direction in the (0,2) theory corresponds
after the  T-duality elementary string 
wound in the  compact direction. This appears as electric flux
 in the worldvolume of the D1-brane.  The composite 
of D-string with electric flux appears as 
 a dyon in the 3+1 SYM.
S-duality in the  $3+1$ Matrix theory therefore corresponds to
transformation 
of momenta from the lightcone direction to the compact direction.
 Scattering of incoming  monopoles 
into outgoing dyons  correspond,  in the $(0,2)$ theory on $S^1$, 
to a state scattering from the lightcone direction 
to the compact direction.

Quantizing zero modes that quantum mechanics 
on monopole moduli space produces the 5 scalars and the 3
vectors fields of the 4+1 SYM vector multiplet.
 This comes from 
the fact that the D3-D1 brane system in IIB has a $Spin(5)_R$ symmetry 
from the transverse space-time.

\subsec{$T^2$ compactification}

By analogy to the  $R^3 \times S^1$ case, 
we expect 
that instantons on $R^2 \times T^2$ 
will  have a Nahm construction which 
is closely related to that for the corresponding
dimensionally reduced theory. Again instantons
on  $R^2 \times T^2$ will be relevant to compactified  
$(0,2)$  in the vacuum sector. 
Reducing the self-dual Yang-Mills equations to $R^2 \times  T^2$, 
we find that the solutions correspond to vortices in 2+1 dimensions.
The equations are $F = D(\phi_1 + \phi_2)$. These equations were 
studied in \Hitchin. The  moduli 
space is  known to be hyper-Kahler, so it will 
admit supersymmetric  quantum mechanics with $8$ supercharges
and $Spin(5)$ R-symmetry.

For the $(40)$ system on the $T^2 \times R^2$ 
we can get some insights  into the moduli space 
by doing a T-duality on the two circles. 
 Now we have D2 branes in $x^0,x^1,x^2$
intersecting D2 branes in $x^0,x^3,x^4$. A D2 brane looks like 
a vortex on the  orthogonal D2 brane.
 This is a singular configurations. 
The corresponding Laplacian has  a logarithmic singularity. 
A similar situation was encountered in \bending
with D4 branes ending on NS 5-branes. 
There the singularity was resolved by going to M-theory where 
the D4 branes become just one M5-brane of topology 
$R^4 \times \Sigma$, where $\Sigma $ is a Riemann surface.  
We expect, in analogy, that  the moduli space 
of solutions will correspond in general  to a smoothed
out intersection of the orthogonal two-branes, 
so that there will be essentially one smooth two-brane
which is a  Riemann 
surface equipped with a holomorphic  mapping to $R^2 \times T^2$. 
The mathematical formulation of this correspondence
between Riemann surfaces and  the instantons
could proceed using spectral surfaces 
\fmw. Some related points are developed in \gukov.

These configurations are relevant to  (0,2) theory 
compactified on a $T^2$. 
For large $T^2$ we expect to have
quantum mechanics on these spaces. 
For small $T^2$ the theory flows to 
$3+1$ Yang Mills. 
where the complex  coupling constant
is given by the complex structure of the 
torus \witcom. 
In this limit we need to consider a  
$2+1$ dimensional   dynamics. 
The $3+1$ dimensional Yang Mills
may be thought as the theory 
on the worldvolume 
of IIB three branes.

One way to describe the dynamics of this system 
is in terms   of a  theory on a two-torus, 
which is obtained by T-dualizing the world-volume 
theory of the 0-branes. Another way would be 
to consider the worldvolume of the 2-brane orthogonal 
to the torus. From the point of view of the latter  theory, 
strings wrapped along cycles of  
the $T^2$  are electrically charged particles.

If we compactify the 
$(0,2)$ theory on a torus and take the limit of  zero area,
we  eliminate  any scale from the 
theory. This leads to the conformally invariant 
$3+1$ SYM.  
 However, the 3+1 SYM is still interacting for appropriate choice 
of complex  structure of the torus.
In the   $2+1$ dimensional Matrix description, we expect that
$2+1$ dimensional theory to flow to an interacting fixed point since 
it  describes an  interacting 
 conformal field theory in the limit that the volume of the 
$T^2 \rightarrow \infty$.
We can contrast this to the 4+1 SYM. In that case, 
the limit of zero
 compactification  radius ,  the theory 
flows to a  non-interacting fixed point, since the coupling constant
is proportional to $R_4$.  

Lorentz invariance of  the  $(0,2)$ theory 
should lead to an interesting symmetry when this 
system is formulated in terms  of the 
world-volume of  the 2-brane orthogonal to the
directions of T-duality. 
An interesting  property of these   $2+1$ Matrix theories 
is that they should have a symmetry mixing 
vortices, and two kinds of W-bosons.
The vortices, we have seen,  are related 
to momentum in the $11$ direction. Momenta in the 
directions $4$ and $5$ become elementary strings 
ending on the two-branes. Viewed from the 2-brane
worldvolume 
orthogonal to the directions of T-duality, the latter 
are W-bosons.  So there is a symmetry exchanging
electrically charged objects with  vortices.

The theory on this two-brane has an $SO(6)$ R-symmetry
from the transverse spacetime, whereas the 4-brane 
worldvolume  theory had an $Spin(5)$ symmetry. 
By quantizing the zero modes of the quantum mechanics on 
the moduli space of vortices in an $SO(6)$ 
covariant fashion, along the lines of 
section 4,  we produce as bosonic states,  6 scalars and 
2 components of a vector field as appropriate for  
 $N=4$ SYM vector multiplet.

\subsec{ $T^4$ compactification.}

In the $R^4$ case we have dealt with quantum mechanics
on a space which is strictly a symmetric product. 
In the case of $T^4$ the appropriate moduli 
spaces for $1$ 4-brane and $N$ zero branes 
 are birational to symmetric products \vainst.     
We can try to  use the  prescription we used to obtain the 
Hilbert space in the case $k=1$ on $R^4$ 
in the case of $T^4$. It is less well-motivated
because the moduli spaces are not 
strictly symmetric products. But it  seems  
reasonable if we believe that the single 5-brane 
continues to behave like   a standard free tensor multiplet
theory when the directions transverse to the light-cone 
are compactified on a torus. After a boost, the system 
is related to 5-brane of M theory  compactified on $T^5$. 
The system of 4-brane, zero-brane and momentum
in IIA theory  is U-dual to 4-brane with momentum in two 
different directions within the 4-brane 
world-volume. This is indeed  counted by a free
field theory because the 4-brane worldvolume theory 
is a $U(1)$ gauge theory.  This means that the prescription 
 we used for the counting of states is consistent with 
the correct number of BPS states, even for states that 
are not ground states in the quantum mechanics.  

In the above we have only discussed a simple class
of states namely 4-branes with 0-branes and momenta.
This is the sector of interest here because,
after decompactification, these states are related 
 to the local operators for  the $(0,2)$ theory 
in the simplest background $R^4 \times R^+ \times R^-$, 
without extended objects of infinite energy.   
It will be interesting to extend this discussion
to systems  with more charges, e.g.  4-brane, 2-brane
and 0-brane, and compare U-duality predictions  with the 
$l_p \rightarrow 0$ limit of  the 5-brane actions
of  \schper\appsch\pt , and to understand the relation 
with the approach of \dvvfve.

{\bf NOTE ADDED }: A previous version of this sub-section 
reported an inconsistency between the picture 
of the single 5-brane as a free field theory, and U-duality. 
This was based on an inaccuracy in tracking the 
appropriate transformations under a sequence of dualities. 
As a result the above discussion  on $T^4$ compactification has been 
rewritten. Our  conclusion on this issue  
 is that there is no conflict between 
U-duality,  the statement that the single M5-brane is a free
 field theory, and the conjecture that M5-branes are described 
 by quantum mechanics on instanton moduli spaces.

\newsec{Summary and comments.} 
 We have tested the Matrix model proposal 
 for a single 5-brane and for multi-5-branes in $R^4$.
 The summary of our tests is in the introduction. 
  The conjecture works well for $R^4$, to the extent that we have 
tested it.  We then discussed  toroidal compactifications 
of the $(0,2)$ theory, presenting a conjecture and some evidence,  
 that it is still described by quantum mechanics on instanton 
moduli spaces when the compactification scale is large
compared to the scales in any correlation function of interest.

 \bigbreak\bigskip\bigskip
\centerline{\bf Acknowledgments}\nobreak
It is a pleasure to thank  O. Ganor, Z. Guralnik, K. Intriligator, 
D.Lowe, J. Park,   S. Sethi, D. Waldram, W. Taylor 
 for discussions. This  work was supported 
by NSF Grant PHY96-00258 and DOE Grant DE-FGO2-91-ER40671.

\listrefs
\end